\documentclass{article}

\usepackage{graphicx}
\usepackage{physics}
\usepackage{amsmath}
\usepackage{amsfonts}
\usepackage{hyperref}
\usepackage{epstopdf}
\usepackage{amssymb}
\usepackage{float}
\usepackage{cite}
\usepackage[noblocks]{authblk}
\usepackage[labelfont=bf,labelsep=none]{caption}
\title{Detuning-Controlled Phase Transition from Passive to Active Regimes in Non-Markovian Quantum Batteries}
\author[]{Meysam Helmi Barati Farimani and Ali Mortezapour \footnote{mortezapour@guilan.ac.ir}}
\affil[]{{\small Department of Physics, University of Guilan, Rasht 41335-1914, Iran}}

\date{}

\begin{document}
\maketitle
\begin{abstract} 
	We investigate a two-qubit quantum battery where coherent charger–battery coupling competes with non-Markovian environmental interactions. By tuning the coupling strengths and detuning, we identify regimes in which environmental memory enhances energy storage and charging power, while strong dissipation suppresses ergotropy by driving the battery into passive states. We show that detuning plays a dual role: reducing dissipation and inducing a phase shift in the memory kernel that controls the interference between coherent energy exchange and environment-induced backflow. As a result, although the stored energy varies smoothly, the extractable work exhibits a discontinuous onset at a critical detuning, signaling a first-order phase transition in ergotropy. The corresponding phase diagram in the coupling–detuning plane reveals a sharp boundary between thermodynamically inactive and work-producing regimes. Our results demonstrate that phase-controlled coherence and non-Markovianity provide a powerful mechanism for optimizing work extraction in open quantum batteries, offering practical strategies for noise-resilient quantum energy storage.
\end{abstract}

\section{Introduction} \label{I}
The exploitation of uniquely quantum mechanical features—such as superposition, entanglement, and coherence—has led to the emergence of quantum technologies capable of outperforming their classical counterparts. Significant advances have been achieved in quantum-enhanced metrology \cite{Vittorio2004quantmeas, Giovannetti2011QuantumMetrology, len2022quantum, yin2023experimental, demkowicz2012elusive}, quantum communication \cite{gisin2007quantum, dominguez2026satellite, clark2025coexistence, ruskuc2025multiplexed}, and quantum information processing \cite{nielsen2010quantum, Suraj2025quantinf}. As experimental platforms continue to mature, extending quantum advantages to thermodynamic and energetic tasks has become a central objective of contemporary research.

Within this broader framework, the concept of the quantum battery (QB) was introduced as a quantum system capable of storing and delivering energy through controlled coherent processes \cite{campaioli2019quantum, binder2015quantacell, Alicki2013entangel, campaioli2017enhancing, ferraro2018high, andolina2019extractable, farina2019charger}. Unlike classical batteries, QBs operate in regimes where quantum correlations and coherence may directly influence charging dynamics and work extraction. Early theoretical investigations demonstrated that collective quantum effects can enhance charging power and produce advantageous scaling with system size \cite{campaioli2017enhancing, ferraro2018high, andolina2019extractable, Mayo2022colleffect, carrasco2022collective, Rossini2020quantadv}. Subsequent studies explored many-body realizations and spin-chain architectures, further clarifying the role of collective interactions in boosting charging performance \cite{PhysRevA.97.022106, rossini2019many, Li2025collcharge, sp5l-c6m8}.

The theoretical description of quantum batteries is embedded in the broader framework of quantum thermodynamics, which extends classical notions of work, heat, and entropy to microscopic quantum systems \cite{gemmer2009quantum, Vinjanampathy01102016, Muller2018corr}. A central concept in this context is ergotropy, defined as the maximum extractable work from a quantum state via cyclic unitary operations \cite{allahverdyan2004maximal, Alicki2013entangel, Yang2023batcap, Gyhm2022quantumcharging, Quach2020darkstate, Song2024remotecharging}. Further investigations into the roles of coherence, correlations, and entanglement in quantum work extraction have clarified their distinct contributions to charging performance and ergotropy. In particular, quantum coherence and correlations can significantly enhance charging power and ergotropy \cite{Simon2025corr, Francica2017DaemonicErgotropy, Perarnau2015extraccorr, OULARABI2025131003}, whereas entanglement is not a strict prerequisite for achieving optimal ergotropy \cite{ Hovhannisyan2013ent}. Nevertheless, under certain collective charging protocols, entanglement may provide an additional advantage by accelerating the charging dynamics.

In realistic implementations, quantum batteries cannot be considered isolated systems and must be described within the theory of open quantum systems, so it is crucial to investigate open quantum batteries \cite{Barra2019disspative, Gherardini2020stab, Zhang2019harmoniccharging, Bai2020floquet, hadipour2025amplified}. Environmental interactions typically induce decoherence and dissipation, leading to degradation of stored energy and reduced charging efficiency \cite{Morrone2023deamonergo}. Several studies have explicitly analyzed dissipative effects on quantum battery performance under Markovian noise, demonstrating reduced charging power, instability, and energy leakage \cite{carrega2020dissipative, Liu2019lossfree}. Barra's pioneering work established the framework for dissipative charging in the Born-Markov regime \cite{Barra2019disspative}, showing that environmental coupling can enable charging even in the absence of coherent chargers, though at the cost of reduced ergotropy.

When environmental memory effects are present, however, the system dynamics deviate from the Markovian approximation. Non-Markovianity which characterized by information and energy backflow from the environment to the system, has been extensively studied and quantified in quantum dynamical maps \cite{rivas2014quantum, Breuer2009measure, bylicka2014non, Bhattacharya_2021}. The characterization of non-Markovianity via information backflow \cite{rivas2014quantum} underlies the identification of environment-assisted charging regimes. In recent years, attention has turned toward the impact of non-Markovian environments on quantum battery dynamics. It has been shown that structured reservoirs can enhance energy storage and charging performance under suitable conditions \cite{OULARABI2025131003, kamin2020non, li2022quantum, Xu2021enhance, Lu2025topqb}. Further investigations revealed that environmental memory may suppress self-discharging and stabilize stored energy \cite{kamin2020non, Bhanja2024impact, xu2024inhibiting}. Beyond environmental engineering, non-Markovianity itself has been identified as a resource for enhanced charging \cite{Bhatt2022nonres}. Strong system-bath coupling beyond perturbative limits requires a non-Markovian treatment \cite{Chin2013quantmet}, which is relevant in moderate-to-strong coupling regimes accessible in current experiments. Experimental control of non-Markovianity via photonic environments \cite{Liu2011expcont} enables direct testing of these predictions. More recent works demonstrated that modulation of bath spectral properties and reservoir engineering can optimize ergotropy, improve charging power, and prolong energy retention in open quantum batteries \cite{xu2024inhibiting, Tabesh2020env, li2025optimal}. These results suggest that non-Markovianity may serve not merely as a detrimental effect but as a controllable resource in quantum energy storage.

Parallel to these developments, the study of driven-dissipative phase transitions has emerged as a unifying framework for nonequilibrium quantum systems \cite{Carmi2015break, Buca2012note, Minganti2018spect}. Such transitions have been extensively studied in optical bistability \cite{Carmi2015break}, many-body condensation \cite{Buca2012note}, and Liouvillian spectral degeneracies \cite{Minganti2018spect}, typically manifesting in steady-state observables or correlation functions. Phase transitions in open quantum systems have primarily been classified according to eigenvalue degeneracies of the Liouvillian superoperator \cite{Minganti2018spect}, whereas their manifestation in fundamentally thermodynamic quantities, particularly in the extractable work, remains unexplored. Meanwhile, investigations of charging transitions in quantum batteries have focused on continuous crossovers in collective systems \cite{De2018reconc, Per2021workext}, leaving open the question of whether genuine phase transitions, particularly first-order discontinuities, can occur in the extractable work of finite open systems. Finite-size systems can exhibit sharp crossover behavior identifiable as phase transitions when appropriate order parameters are considered \cite{Zan2006ground}, analogous to magnetization in quantum spin systems \cite{Sachdev2011quant}.

Despite these significant advances, the combined influence of coherent charger–battery interaction and independent non-Markovian environments coupled simultaneously to both subsystems remains insufficiently explored. In particular, a systematic analysis of how detuning, coupling strengths, and reservoir memory jointly influence energy variation, charging power, and ergotropy is still lacking. Previous studies of charging transitions \cite{De2018reconc, Per2021workext} considered either closed systems or Markovian environments, where the phase structure differs qualitatively from that in structured reservoirs.

Motivated by this gap, we investigate a quantum battery model in which the charger and battery interact coherently while each subsystem simultaneously couples to its own structured non-Markovian environment. This configuration generates rich dynamical behavior arising from the competition between coherent energy exchange and environment-induced memory effects. By tuning the coupling strengths, detuning parameters, and environmental spectral characteristics, we identify operational regimes that optimize energy storage, enhance charging power, and maximize ergotropy under realistic open-system conditions. Most notably, we discover that detuning induces a first-order phase transition in the extractable work: while stored energy varies smoothly, ergotropy exhibits a discontinuous onset at a critical detuning value, signaling a sharp boundary between thermodynamically passive and active regimes. This discontinuity, hidden in smooth energy observables, reveals that work extraction is fundamentally a phase-sensitive phenomenon in quantum systems.

The remainder of this paper is organized as follows: Section \ref{II} introduces the principal performance metrics, such as energy variation, average charging power, and ergotropy, and then discusses their physical interpretation. Section \ref{III} presents the theoretical model and dynamical equations. Section \ref{IV} contains a detailed analysis of the results. Section \ref{V} briefly discusses the feasibility of experimental realization. Finally, we summarize our findings in Section \ref{VI}.

\section{Energy variation, average charging power and extractable work} \label{II}
A quantum battery (QB) is a device that can store energy. The process of storing energy in a QB is governed by the principles of quantum mechanics. A QB is considered efficient if it can store energy in the shortest possible time and retrieve the maximum amount of energy when needed. There are three important metrics to analyze the performance of a QB: energy variation, which is the amount of energy that can be stored; average power for charging; and the maximum amount of work that can be extracted from the QB during a cyclic process.

The energy stored in a QB at time $t$ is given by
\begin{equation}\label{eq1}
	\Delta E_B(t) = \text{Tr}\left[\hat{H}_B \hat{\rho}_B(t)\right] - \text{Tr}\left[\hat{H}_B \hat{\rho}_B(0)\right],
\end{equation}
where $\rho_B$ and $H_B$ are the density matrix and free Hamiltonian of the QB, respectively. In order to quantify the average power for charging a QB at time $t$, we define the following expression:
\begin{equation}\label{eq2}
	P_B(t) = \frac{E_B (t)}{t}.
\end{equation}

Additionally, an important concept in the field of quantum thermodynamics is $ergotropy$, which is the maximum amount of work that can be extracted from a QB during a cyclic process \cite{allahverdyan2004maximal}:
\begin{equation}\label{eq3}
	W = \text{Tr} \left[\hat{H}_B \hat{\rho}_B\right] - \min_U \text{Tr} \left[\hat{H}_B \hat{U} \hat{\rho}_B \hat{U}^\dagger\right],
\end{equation}
where minimization is performed over all achievable unitary transformations. For any specified $\rho_B$, there exists a singular state that optimizes this relation, known as the passive state. Thus, ergotropy can be rewritten as:
\begin{equation}\label{eq4}
	W = \text{Tr} \left[\hat{H}_B \hat{\rho}_B\right] - \text{Tr} \left[\hat{H}_B \hat{\sigma}_{\rho_B}\right],
\end{equation}
where $\sigma_{\rho_B}$ is the passive state associated with $\rho_B$. It is worth mentioning that work cannot be extracted from a QB if its state is a passive state \cite{allahverdyan2004maximal}.

Using spectral decomposition, one can easily express the density matrix $\rho_B$ and its associated Hamiltonian $H_B$ in the following form:
\begin{align}\label{eq5}
	\hat{H}_B & = \sum_{i} \epsilon_i \ket{\epsilon_i} \bra{\epsilon_i}, \quad \epsilon_1 \leq \epsilon_2 \leq \cdots \leq \epsilon_i, \\
	\hat{\rho}_B & = \sum_{j} r_j \ket{r_j} \bra{r_j}, \quad r_1 \geq r_2 \geq \cdots \geq r_j. \notag
\end{align} 
Here, $\epsilon_i$ ($\ket{\epsilon_i}$) and $r_j$ ($\ket{r_j}$) are the eigenvalues (eigenstates) of the Hamiltonian and density matrix, respectively.

Thus, the passive state can be rewritten as \cite{allahverdyan2004maximal}:
\begin{equation}\label{eq6}
	\hat{\sigma}_{\rho_B} = \hat{U} \hat{\rho}_B \hat{U}^\dagger = \sum_{i} r_i \ketbra{\epsilon_i}{\epsilon_i}.
\end{equation}
By substituting Eqs.~\eqref{eq5} and \eqref{eq6} into Eq.~\eqref{eq4}, the expression for ergotropy takes the following form \cite{allahverdyan2004maximal}:
\begin{equation}\label{eq7}
	W = \sum_{ij} r_j \epsilon_i \left(\abs{\braket{r_j}{\epsilon_i}}^2 - \delta_{i,j}\right).
\end{equation}
Here, $\delta_{i,j}$ is the Kronecker delta function.
\section{Model} \label{III}
\begin{figure}[ht]
	\centering
	\includegraphics[width= 1\linewidth]{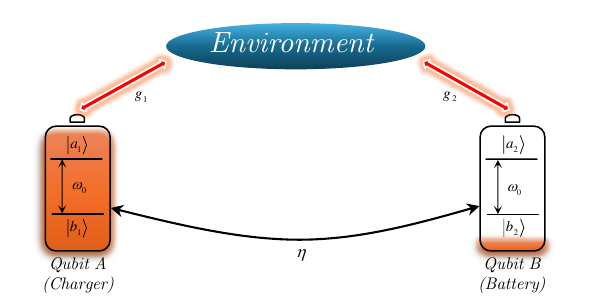}
	\caption{Schematic illustration of the charging process (the considered model). A quantum charger (A) coherently interacts with a quantum battery (B) with a coupling rate $\eta$. The charger and the battery are coupled to a common environment with different coupling strengths.}
	\label{fig1}
\end{figure}
The model under consideration comprises the charger (qubit A) and the QB (qubit B). Each qubit has excited and ground states, represented as $\ket{e_i}$ and $\ket{g_i}$, respectively. The quantum charger coherently interacts with QB with a coupling rate $\eta$. Both charger and battery are embedded in a non-Markovian zero-temperature environment and coupled asymmetrically to the quantized modes of the environment. Fig.~\ref{fig1} illustrates a schematic representation of the model. The Hamiltonian for the total system 
$(\hat{H})$ is divided into three components: the two-qubit Hamiltonian (denoted as 
$\hat{H}_S$), the Hamiltonian that describes the evolution of the environment (denoted as 
$\hat{H}_E$),
and the Hamiltonian representing  the interaction between environment and each qubit (denoted as 
$\hat{H}_I$). The total Hamiltonian of the system can be written as
\begin{equation}\label{eq8}
	\hat{H} = \hat{H}_S + \hat{H}_E + \hat{H}_I
\end{equation}
Where
\begin{align}
	\begin{split}\label{eq9}
		\hat{H}_S &= \omega_0 \sum_{i=1,2} \sigma^{+}_i \sigma^{-}_i
		 + \frac{\eta}{2} (\sigma^{+}_1 \sigma^{-}_2 + \sigma^{-}_1 \sigma^{+}_2),\\
		\hat{H}_E&=\sum_{k} \omega_k \hat{a}_k^{\dagger} \hat{a}_k,\\
		\hat{H}_I&=\sum_{i=1,2} \sum_{k} g_i\left(\omega_k\right) \left(\hat{a}_k \sigma_i^{+} + \hat{a}_k^{\dagger} \sigma_i^{-} \right),
	\end{split}
\end{align}
where $\omega_0$ represent the transition frequency between excited state $\ket{a}$ and ground state $\ket{b}$ of each qubit. The variable $\eta$ denotes the strength of the interaction  between the qubits, while $g_i \left(\omega_k\right) \left(i=1,2\right)$ denotes the coupling strength between qubits and the environment. The creation and annihilation operators for $k$-th mode of the bosonic environment are labeled as $\hat{a}_k^{\dagger}$ and $\hat{a}_k$, respectively. Additionally, the rising and lowering operators for the qubits are defined as 
$\sigma_i^+=\ketbra{a_i}{b_i}$ and
$\sigma_i^-=\ketbra{b_i}{a_i}$, with $i=1,2$.

We assume the initial state of the total system is considered to be as follows:
\begin{align}\label{eq10}
	\begin{split}
		&\ket{\Psi(0)}=\left(C_1(0) \ket{a_1} \ket{b_2} + C_2(0) \ket{b_1} \ket{a_2}\right) \ket{0}_E,\\
		&C_1(0)= \cos(\theta/2),\\
		&C_2(0) = \sin(\theta/2) e^{i \phi},
	\end{split}
\end{align}
in which $\ket{0}_E$ characterizes the vacuum state of the reservoir. Hence, the time evolution of the total system, under the action of the effective Hamiltonian (Eq.~\eqref{eq8}), is described by
\begin{equation}\label{eq11}
	\ket{\Psi(t)}=C_1(t) \ket{a_1,b_2}\ket{0}_E + C_2(t) \ket{b_1,a_2}\ket{0}_E+ \sum_{k} C_{b,k} (t) \ket{b_1,b_2}\ket{1_k},
\end{equation}
where $\ket{1_k}$ signifies the state of the reservoir possessing one photon in the $k$-th vacuum mode. By applying the amplitude transformations 
$C_i(t)=\tilde{C_i} (t) e^{-i\omega_0 t}$ and
$C_{b,k}(t) = \tilde{C}_{b,k}(t) e^{-i\omega_k t}$, and substituting Eq.~\eqref{eq11} into the Schrödinger equation, one can derive the equations of motion for the tilde amplitudes as follows:
\begin{align}
	& \frac{\dd{\tilde{C}}_1}{\dd{t}} = -i \frac{\eta}{2} \tilde{C}_2 (t) - \sum_{k} g_1 \left(\omega_k\right) e^{-i\delta_k t} \tilde{C}_{b,k}(t), \label{eq12}\\
	& \frac{\dd{\tilde{C}}_2}{\dd{t}} = -i \frac{\eta}{2} \tilde{C}_1 (t) - \sum_{k} g_2 \left(\omega_k\right) e^{-i\delta_k t} \tilde{C}_{b,k}(t),\label{eq13}\\
	&\frac{\dd{\tilde{C}}_{b,k}}{\dd{t}} =-i \left(\tilde{C}_1 (t) g_1 \left(\omega_k\right) - \tilde{C}_1 (t) g_2 \left(\omega_k\right)\right)e^{-i\delta_k t}\label{eq14}
\end{align}
where $\delta_k = \omega_k - \omega_0$. By assuming $\tilde{C}_{b,k} (0) = 0$ and taking formal integral of Eq. \eqref{eq14} and successively substituting it into Eqs. \eqref{eq12} and \eqref{eq13} results in:
\begin{align}
	& \frac{\dd{\tilde{C}_1 (t)}}{\dd{t}} = -i \frac{\eta}{2} \tilde{C}_2(t) - \int_{0}^{t} \sum_{k} e^{-i\delta_k(t-t^\prime)} \left[\abs{g_1}^2 \tilde{C}_1 (t^\prime) - g_1 g_2 \tilde{C}_2 (t^\prime)\right] \dd{t^\prime},\label{eq15}\\
	& \frac{\dd{\tilde{C}_2 (t)}}{\dd{t}} = -i \frac{\eta}{2} \tilde{C}_1(t) - \int_{0}^{t} \sum_{k} e^{-i\delta_k(t-t^\prime)} \left[\abs{g_2}^2 \tilde{C}_2 (t^\prime) - g_1 g_2 \tilde{C}_1 (t^\prime)\right]\dd{t^\prime},\label{eq16}
\end{align}
In the continuum limit for reservoir spectrum, the summation over $k$ is replaced by an integration according 
$\sum_{k} \rightarrow \int J(\omega) \dd{\omega}$, where $J(\omega)$ is the spectral density of the cavity modes, defines using the following Lorentzian form:
\begin{equation}\label{eq17}
	J(\omega) = \frac{1}{\pi} \frac{\gamma/2}{(\omega - \omega_c)^2 + (\gamma/2)},
\end{equation}
where $\gamma$ and $\omega_c$ are the width and the peak of the distribution, respectively. If the coupling strengths $g_i(\omega_k)$ for $i=1,2$, change slowly as function of $\omega_k$ around $\omega_c$, we can substitute  these strengths with their values at this frequency for a good approximation. In this case, we define $g_i \equiv g_i(\omega_c)$ for $i=1,2$ to simplify the notation. The resulting two integrodiffrential equations for $\tilde{C}_1 (t)$ and $\tilde{C}_2 (t)$ then are
\begin{align}
	& \frac{\dd{\tilde{C}_1 (t)}}{\dd{t}} = -i \frac{\eta}{2} \tilde{C}_2(t) - \int_{0}^{t} f(t-t^\prime) \left[g_1^2 \tilde{C}_1 (t^\prime) - g_1 g_2 \tilde{C}_2 (t^\prime)\right] \dd{t^\prime},\label{eq18}\\
	& \frac{\dd{\tilde{C}_2 (t)}}{\dd{t}} = -i \frac{\eta}{2} \tilde{C}_1(t) - \int_{0}^{t} f(t-t^\prime) \left[g_2^2 \tilde{C}_2 (t^\prime) - g_1 g_2 \tilde{C}_1 (t^\prime)\right]\dd{t^\prime},\label{eq19}
\end{align}
in which the kernel $f(t-t^\prime)$ is the correlation function ans is defined as
\begin{equation}\label{eq20}
	f(t-t^\prime) = \int_{- \infty}^{\infty} J(\omega) e^{-i\delta_k (t-t^\prime)} \dd{\omega} =\frac{\gamma}{2 \pi} \int_{- \infty}^{\infty} \frac{e^{-i(\omega - \omega_{0})}\dd{\omega}}{(\omega - \omega_c)^2 + (\gamma/2)} = e^{-(\gamma/2 + i\Delta) (t-t^\prime)},
\end{equation}
where $\Delta = \omega_c - \omega_{0}$. Regarding Eq.~\eqref{eq20} and taking the Laplace transform of Eqs.~\eqref{eq18} and \eqref{eq19}, one obtains:
\begin{align}
	& s F_1 (s) =C_1 (0) - i \frac{\eta}{2} F_2 (s) - g_1^2 \Lambda (s) F_1 (s)- g_1 g_2 \Lambda (s) F_2 (s), \label{eq21}\\
	& s F_2 (s) = C_2 (0) - i \frac{\eta}{2} F_1 (s) - g_2^2 \Lambda (s) F_2 (s)- g_1 g_2 \Lambda (s) F_1 (s), \label{eq22}
\end{align}
where $F_1 (s)$ and $F_2 (s)$ are Laplace transform of the amplitudes $\tilde{C}_1 (t)$ and $\tilde{C}_2 (t)$ respectively, while 
$\Lambda (s) = 1/ (i\Delta + s + \gamma/2)$.

Once $\tilde{C}_1 (t)$ and $\tilde{C}_2 (t)$ are calculated, the evolved reduced density matrix of two-qubit system $\rho(t)$ in the basis 
$\left\{\ket{a_1,a_2}, \ket{a_1,b_2},\ket{b_1,a_2}, \ket{b_1,b_2}\right\}$
can be acquired as
\begin{equation}\label{eq23}
	\rho(t)=
	\begin{pmatrix}
		0 &0 &0 &0 \\
		0 & \rho_{22}(t) & \rho_{23}(t) &0 \\
		0 & \rho_{32}(t) &\rho_{33}(t) &0 \\
		0 &0 &0 & \rho_{44}(t)
	\end{pmatrix}
	.
\end{equation}
The matrix elements are calculated as follow:
\begin{align}\label{eq24}
	\begin{split}
		&\rho_{22}(t)=\abs{C_1 (t)}^2 = \abs{\tilde{C}_1 (t)}^2,\\
		&\rho_{33}(t)=\abs{C_2 (t)}^2 = \abs{\tilde{C}_2 (t)}^2,\\
		&\rho_{44}(t)=1-\rho_{22}(t)-\rho_{33}(t),\\
		&\rho_{23}(t) = \rho_{32}^\ast(t) = C_1 (t) C_2^{\ast} (t)=\tilde{C}_1 (t) \tilde{C}_2^{\ast} (t).
	\end{split}
\end{align}
Also, the eigenvalues of this density matrix are obtained as follow:
\begin{align}\label{eq25}
	\begin{split}
		& \lambda_1 = 0,\\
		& \lambda_{2,3} = (\rho_{22}(t) + \rho_{33}(t) \pm \sqrt{(\rho_{22}(t) - \rho_{33}(t))^2 + 4 \abs{\rho_{23}(t)}^2}\ )/2,\\
		& \lambda_4 = \rho_{44}(t).
	\end{split}
\end{align}
 
\section{Results and discussion} \label{IV}
In this section, we analyze the performance of the QB based on the three mentioned metrics: energy variation, ergotropy, and average charging power. The reduced density matrix at a specific time $t$ for the charger and QB can be easily obtained by taking a partial trace over A and B in Eq.~\eqref{eq23} \cite{Tabesh2020env},
\begin{align}\label{eq26}
	\begin{split}
		\rho_A (t) &= \abs{C_1 (t)}^2 \ketbra{a_1}{a_1} + \left(1-\abs{C_1 (t)}^2\right) \ketbra{b_1}{b_1}, \\
		\rho_B (t) &= \abs{C_2 (t)}^2 \ketbra{a_2}{a_2} + \left(1-\abs{C_2 (t)}^2\right) \ketbra{b_2}{b_2}.
	\end{split} 
\end{align}
By substituting Eq.~\eqref{eq26} and the free Hamiltonian of the QB, $\hat{H}_B = (\omega_{0} /2) \hat{\sigma}_z$, into Eq.~\eqref{eq1}, the energy variation of the QB at time $t$ can be derived as:
\begin{equation}\label{eq27}
	\Delta E_B (t) = W_0 \left(\abs{C_2 (t)}^2 - \abs{C_2 (0)}^2\right).
\end{equation}
Similarly, ergotropy can be easily derived as follows:
\begin{equation}\label{eq28}
	W = W_0 \left[2 \abs{C_2 (t)}^2 - 1\right] \Theta \left(\abs{C_2 (t)}^2 - \frac{1}{2}\right),
\end{equation}
where $W_0 =\hbar \omega_{0}$ (See \cite{Tabesh2020env}) and $\Theta(x)$ is the Heaviside step function .

We assume that the charger initially possesses the maximum amount of energy, while the QB is completely empty. Thus, the initial state of the whole system is given by:
\begin{equation}\label{eq29}
	\ket{\psi(0)} = \ket{a_1, b_2} \otimes \ket{0}_E,
\end{equation}
which implies that $C_1(0) = 1$ and $C_2(0) = 0$.

In this section, we analyze the charging dynamics of the quantum battery (QB) and clarify the physical mechanisms behind the observed energy variation, ergotropy, and average charging power. For the sake of simplicity, we take $W_0=1$.

\begin{figure}[H]
 	\centering
 	\includegraphics[width=1 \linewidth]{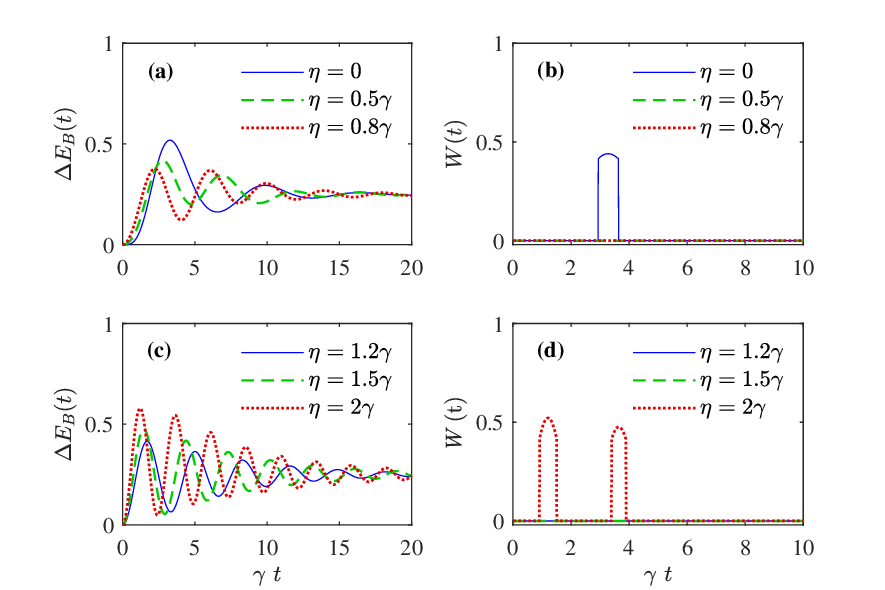}
 	\caption{The energy variation of the QB ((a) and (c)) and ergotropy ((b) and (d)) as a function of the scaled time $\gamma~t$ for different values of coherent coupling $\eta$. Here, $\Delta = 0$ and $g_1 = g_2 = 0.7 \gamma$.}
 	\label{fig2}
 \end{figure}

Fig.~\ref{fig2} illustrates the time evolution of the QB energy variation $\Delta E_B(t)$ and ergotropy $W$ for different values of the charger–battery coupling $\eta$, with $\Delta = 0$ and $g_1 = g_2 = 0.7 \gamma$. Fig.~\ref{fig2} (a,c) show that creating and then increasing the coherent coupling $\eta$ between the charger and the QB enhances oscillations in the stored energy $\Delta E_B(t)$ and delays the approach to steady state. A stronger $\eta$ causes faster Rabi-like energy swapping between the two qubits, allowing the stored energy to flow back from the QB to the charger periodically. In the non-Markovian regime, this coherent exchange combines with environment-induced memory backflow, resulting in prolonged oscillations rather than monotonic relaxation. Note that for $\eta = 0$, the only energy transfer channel is the environment. Here, partial charging happens through environment-assisted excitation exchange, and the system rapidly settles into a stable energy plateau. As $\eta$ increases, the coherent pathway competes with dissipation: oscillatory exchange slows the effective energy loss to the bath, extending the charging period and increasing transient energy peaks.
Fig.~\ref{fig2}(b,d) depict ergotropy $W$, which indicates stored work-usable energy. For intermediate $\eta$ values (0.5–1.5), coherence-driven oscillations trap population in mixed non-passive states that still do not surpass the ergotropy threshold. Only when $\eta = 0$ and $\eta = 2\gamma$ does the coherent energy exchange push the QB into a sufficiently population-inverted, ordered state to produce extractable work, with the energy change of the QB exceeding 0.5. Therefore, large $\eta$ both boosts peak energy and enhances the quality of stored energy by encouraging coherent excitation transfer.

\begin{figure}[H]
	\centering
	\includegraphics[width=0.5 \linewidth]{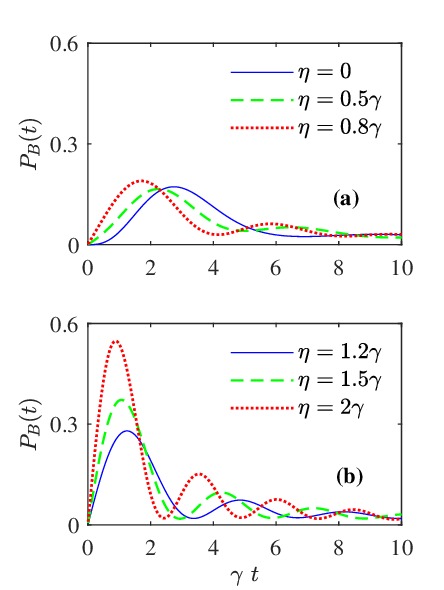}
	\caption{The average charging power of the QB as a function of the scaled time $\gamma~t$ for different values of coherent coupling $\eta$. All parameters are the same as in Fig.~\ref{fig2}.}
	\label{fig3}
\end{figure}

\begin{figure}[H]
	\centering
	\includegraphics[width=1 \linewidth]{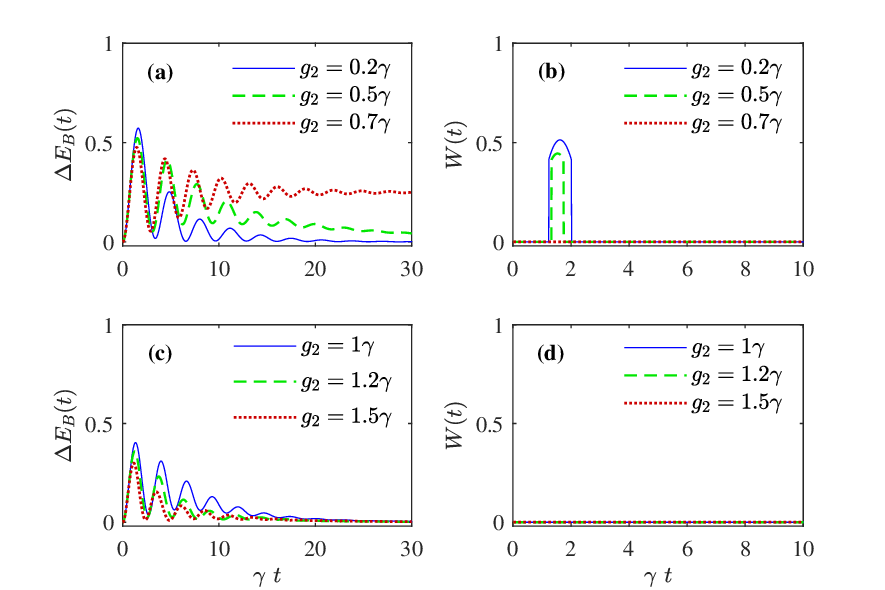}
	\caption{The energy variation of the QB ((a) and (c)) and ergotropy ((b) and (d)) as a function of the scaled time $\gamma~t$ for different values of $g_2$ with $\eta = 1.5 \gamma$, $\Delta = 0$ and $g_1 = 0.7 \gamma$.}
	\label{fig4}
\end{figure}

Fig.~\ref{fig3} shows the average charging power $P_B(t)$ of the QB for different $\eta$. Other parameters are those considered in Fig.~\ref{fig2}. It is seen that increasing $\eta$ results in a marked enhancement of charging power. This behavior reflects the fact that stronger coupling accelerates coherent population transfer from the charger to the QB. The elevated oscillations at higher $\eta$ stem from energy being repeatedly swapped between the two qubits before dissipation acts, effectively boosting instantaneous charging rates. Thus, coherent coupling provides a power-enhancement mechanism beyond purely dissipative charging.
Fig.~\ref{fig4} displays  $\Delta E_B(t)$ and $W$ for different QB–environment coupling $g_B$, with $\eta = 1.5\gamma$,  $\Delta = 0$, and $g_1 = 0.7 \gamma$. For weak to moderate environment coupling ($g_2 \lesssim 0.7\gamma$), the energy variation increases because the structured non-Markovian bath temporarily feeds energy back into the system, effectively supporting environment-assisted charging (Fig.~\ref{fig4}(a)). However, ergotropy gradually decreases as $g_2$ increases (Fig.~\ref{fig4}(b)), indicating that environmental noise disrupts coherence and pushes the QB toward passive states. For strong coupling ($g_2 \geq 0.7\gamma$), the QB becomes highly dissipative. In this case, the energy leaks into the environment faster than it can flow back, suppressing oscillations and lowering the maximum stored energy (Fig.~\ref{fig4}(c)). Therefore, the QB relaxes into a passive state that cannot deliver work (Fig.~\ref{fig4}(d)). This illustrates a transition from beneficial to harmful environmental coupling.

\begin{figure}[H]
	\centering
	\includegraphics[width=0.5 \linewidth]{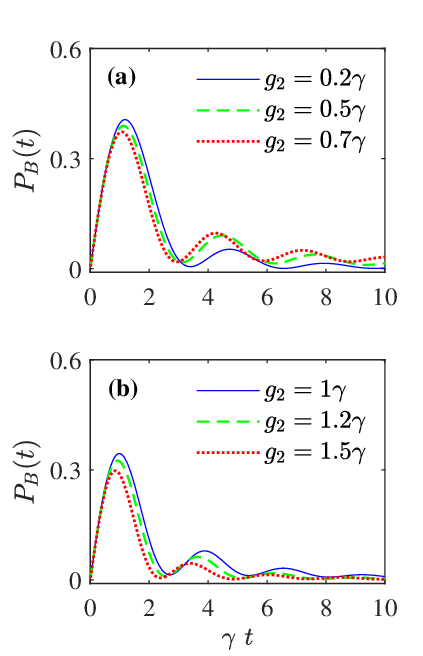}
	\caption{Average charging power of QB as a function of the scaled time $\gamma~t$ for different values of $g_2$. All parameters are the same as in Fig.~\ref{fig4}.}
	\label{fig5}
\end{figure}

Fig.~\ref{fig5} illustrates how increasing $g_2$ affects $P_B(t)$ under the same conditions as Fig.~\ref{fig4}. The charging power steadily decreases as $g_2$ increases. While weak coupling allows for environment-assisted excitation return, stronger coupling leads to rapid loss of population to the reservoir, overpowering both coherent transfer and memory backflow. Consequently, the QB receives energy less efficiently and settles at a lower power level, confirming that strong dissipation reduces performance.

\begin{figure}[H]
	\centering
	\includegraphics[width=0.55 \linewidth]{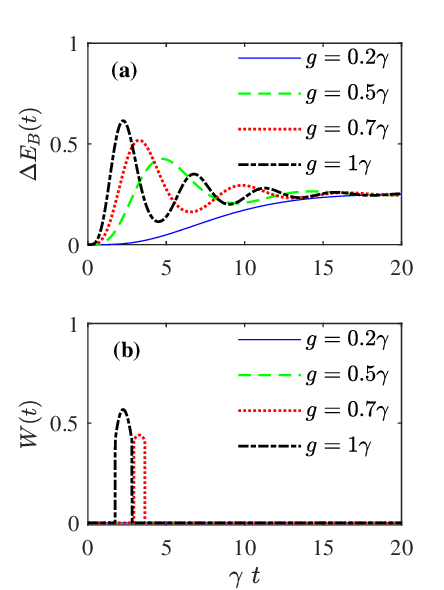}
	\caption{(a) The energy variation of the QB and (b) ergotropy as a function of the scaled time $\gamma~t$ for different values of $g$ with $\eta = \Delta = 0$.
	}
	\label{fig6}
\end{figure}

We next examine the case $\eta = 0$, where the qubits interact with the environment but do not exchange energy directly. In this context, we analyze the effect of $g_1=g_2=g$ on the time evolution of $\Delta E_B(t)$ and $W$ for $\Delta = 0$ in Fig.~\ref{fig6}. It is observed that both energy storage and ergotropy increase with higher $g_i$. Physically, this behavior results from the enhanced memory effect, where the bath returns previously leaked excitation, enabling population inversion. 

\begin{figure}[H]
	\centering
	\includegraphics[width=0.55 \linewidth]{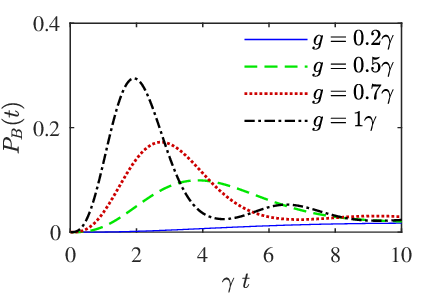}
	\caption{Average charging power as a function of the scaled time $\gamma~t$ for different values of $g$. All parameters are the same as in Fig.~\ref{fig6}.
	}
	\label{fig7}
\end{figure}

Fig.~\ref{fig7} displays $P_B(t)$ under the same conditions as Fig.~\ref{fig6}. The charging power rises with $g$, confirming that reservoir-mediated energy backflow not only increases energy retention but also boosts power delivery. Therefore, strong non-Markovian features enable the environment to serve as an active work resource rather than a loss channel.

\begin{figure}[H]
	\centering
	\includegraphics[width=1 \linewidth]{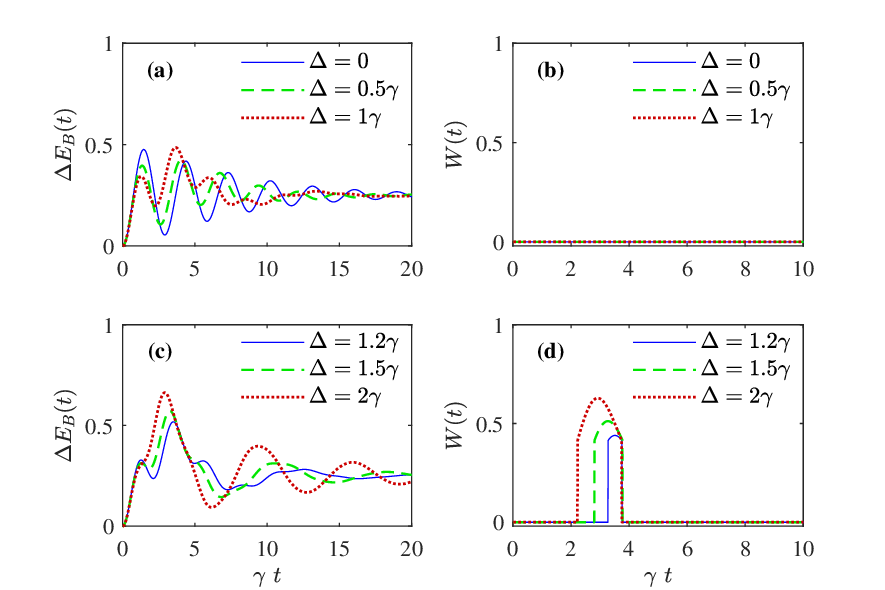}
	\caption{The energy variation of the QB ((a) and (c)) and ergotropy ((b) and (d)) as a function of the scaled time $\gamma~t$ for different values of $\Delta$. Here, $\eta = 1.5\gamma$ and $g_1 = g_2 = 0.7\gamma$.
	}
	\label{fig8}
\end{figure}

We are finally interested in exploring the impact of detuning on battery performance. Fig.~\ref{fig8} exhibits how the energy of the QB and ergotropy change over scaled time for different $\Delta$ values, while keeping $\eta$, $g_A1$, and $g_2$ fixed (with $\eta = 1.5\gamma$ and $g_1 = g_2 = 0.7\gamma$). For $\Delta < 1.2\gamma$, both energy variation and ergotropy decrease because detuning limits resonant energy transfer, reducing both coherent swapping and environment-assisted backflow. In cases where $\Delta \geq 1.2\gamma$, detuning protects the QB from environmental loss by separating system and bath modes energetically. This helps maintain coherence, leading to higher energy oscillations and delayed but stronger ergotropy. The crossover indicates a transition from resonance-limited charging to decoherence-protected charging.

\begin{figure}[H]
	\centering
	\includegraphics[width=0.5 \linewidth]{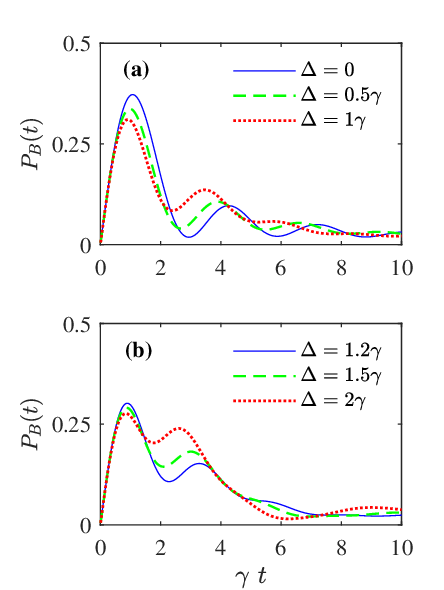}
	\caption{Average charging power as a function of the scaled time $\gamma~t$ for different values of $\Delta$. All parameters are the same as in Fig.~\ref{fig8}.
	}
	\label{fig9}
\end{figure}

Fig.~\ref{fig9} displays $P_B(t)$ for different detuning values. For $\Delta < 1.2\gamma$, charging power decreases as energy transfer becomes inefficient. In cases where $\Delta \geq 1.2\gamma$, $P_B(t)$ increases as detuning suppresses dissipation and preserves coherent dynamics. Thus, tuning $\Delta$ provides a knob to transition between fast-damping resonant charging and coherence-protected high-power charging.

\begin{figure}[H]
	\centering
	\includegraphics[width=0.6 \linewidth]{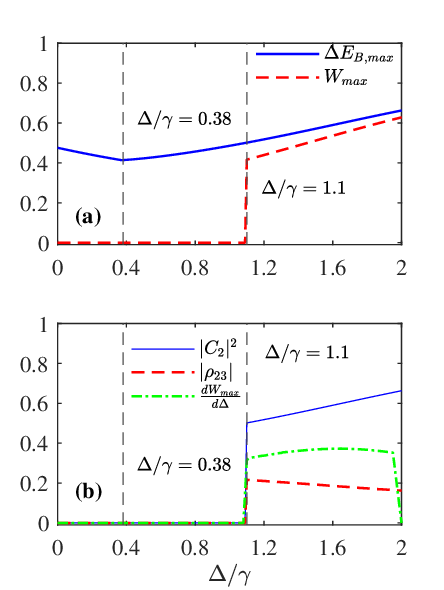}
	\caption{(a) $\Delta E_{B, \text{max}}$ and $W_{\text{max}}$ and (b) $C_{2}$ , $|\rho_{22}|$ and $\dd{W_{\text{max}}} / \dd{\Delta}$ versus $\Delta/\gamma$. The other parameters are taken as: $\eta = 1.5\gamma$ and $g_1 = g_2 = 0.7\gamma$ . 
	}
	\label{fig10}
\end{figure}

Fig.~\ref{fig10} reveals a detuning-induced first-order phase transition in the extractable work. While the maximum stored energy increases smoothly with detuning, ergotropy remains strictly zero up to a critical point $\Delta_{c}=\Delta = 1.1\gamma$, beyond which it exhibits a discontinuous jump. This variation indicates that energy may be accumulated without being extractable until coherence is sufficiently protected. The behavior originates from a detuning-controlled phase shift in the system dynamics. Specifically, the memory kernel introduces a phase factor $e^{-i\Delta t}$, which governs the interference between coherent energy exchange and environment-induced backflow. Below $\Delta_{c}$, destructive interference prevents population inversion, keeping the battery in a passive state. Above $\Delta_{c}$, a critical phase shift induces constructive interference, leading to a sudden enhancement of excitation transfer and the emergence of non-passive states. The discontinuities observed in population, density matrix elements, and the derivative of ergotropy confirm the first-order nature of the transition.

\begin{figure}[H]
	\centering
	\includegraphics[width=0.6 \linewidth]{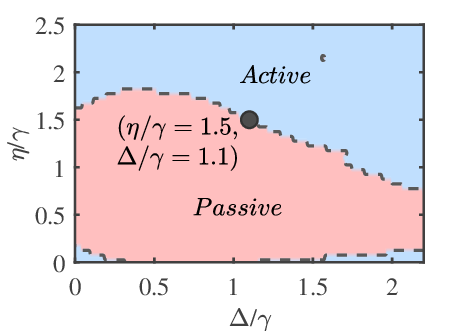}
	\caption{$W_{\text{max}}$ as a function of $\eta/\gamma$ and $\Delta/\gamma$ for $g_1 = g_2 = 0.7\gamma$ .
	}
	\label{fig11}
\end{figure}

Fig.~\ref{fig11} generalizes the behavior observed in Fig.~\ref{fig10} into a phase diagram in the ( $\eta$ , $\Delta$) plane, clearly identifying a boundary between passive and active regimes. This boundary reflects the combined role of coherent coupling strength and detuning-induced phase control, demonstrating that extractable work emerges

\section{Experimental Feasibility} \label{V}
The two-qubit model with independent baths is immediately realizable in several platforms. Superconducting transmon qubits coupled to engineered resonator baths \cite{Hu2025quant} naturally implement the non-Markovian regime when bath correlation times exceed qubit relaxation times. The detuning parameter $\Delta$  corresponds to qubit frequency tuning via external flux bias, while the coupling  $\eta$ is realized through capacitive or inductive qubit-qubit coupling. The asymmetric coupling strengths $g_{1} \neq g_{2}$ can be achieved by geometric arrangement or frequency-selective filtering. Our predicted phase transition would manifest as a sudden onset of microwave amplification when $\Delta$ crosses the critical value, detectable via standard dispersive readout.

\section{Conclusion} \label{VI}

In this work, we have analyzed the charging dynamics of a two-qubit quantum battery subject to both coherent charger–battery interaction and non-Markovian environmental coupling. By systematically exploring the roles of coherent coupling strength, system–bath interaction, and detuning, we have shown that the performance of the battery is governed by a nontrivial interplay between coherent energy exchange, dissipation, and environmental memory effects.

Our results demonstrate that moderate non-Markovian coupling can enhance energy storage and charging power through environment-assisted backflow, while strong dissipation suppresses ergotropy by driving the system toward passive states. Increasing the coherent coupling strength promotes Rabi-like oscillations, enabling faster energy transfer and higher charging power, while also improving the quality of stored energy. These findings align with recent experimental progress in controlling non-Markovian dynamics in quantum-optical platforms \cite{Liu2011expcont}, suggesting the immediate realizability of our predicted effects.

The central result of this work is the identification of a detuning-controlled phase-shift mechanism that fundamentally alters the thermodynamic behavior of the battery. We have shown that detuning not only suppresses environmental losses by energetically separating system and bath modes but also induces a relative phase shift in the system dynamics, modifying the interference between coherent and memory-assisted energy transfer pathways. As a consequence, the system undergoes a transition from a passive energy-storage regime to an active work-extraction regime at a critical detuning.

Most notably, we have demonstrated that this mechanism gives rise to a first-order phase transition in the extractable work. While the maximum stored energy varies smoothly with detuning, the ergotropy exhibits a discontinuous onset at a critical point $\Delta_{c}=1.1 \gamma$, accompanied by abrupt changes in population and non-analytic behavior in its derivative. This discontinuity, hidden in smooth energy observables, reveals that work extraction is fundamentally a phase-sensitive phenomenon in quantum systems. The corresponding phase diagram in the coupling–detuning plane reveals a sharp boundary between thermodynamically inactive and work-producing regimes, demonstrating that extractable work emerges only when both coherent coupling and detuning exceed critical thresholds.

This discovery establishes a new class of thermodynamic critical phenomena in open quantum systems, complementing studies of driven-dissipative phase transitions in optical \cite{Carmi2015break} and many-body systems \cite{Buca2012note, Minganti2018spect} by demonstrating that work extraction itself can serve as an order parameter. Unlike previous investigations of charging transitions in quantum batteries that identified continuous crossovers in collective systems \cite{De2018reconc, Per2021workext}, our first-order discontinuity in a minimal two-qubit system reveals that sharp thermodynamic boundaries can emerge even in finite systems when appropriate order parameters—here, the ergotropy—are considered \cite{Zan2006ground, Sachdev2011quant}. The non-analytic behavior we observe in ergotropy derivatives mirrors critical phenomena in entanglement and non-Markovianity measures \cite{Breuer2009coll}, suggesting universal features of quantum phase transitions in open systems.

Our findings carry significant implications for experimental quantum thermodynamics. The parameters required to observe the predicted phase transition—moderate qubit-qubit coupling $\eta \sim \gamma$, asymmetric bath couplings $g_{1} \neq g_{2}$, and tunable detuning  $\Delta$ —are readily achievable in superconducting circuit architectures \cite{sp5l-c6m8} and cavity-QED systems \cite{Liu2011expcont}. The discontinuous onset of microwave amplification or work extraction at the critical detuning would provide unambiguous signatures of the transition, accessible via standard dispersive readout techniques. Moreover, the phase-shift control mechanism we identify offers a practical knob for optimizing quantum battery performance without requiring precise engineering of bath spectral properties.\\
Several directions merit future investigation. Extending our analysis to finite temperatures would clarify the robustness of the phase transition against thermal fluctuations, relevant for near-term devices operating at millikelvin temperatures. Incorporating quantum correlations and entanglement between charger and battery as resources \cite{ Hovhannisyan2013ent} may reveal additional phase structure in the global versus local work extraction. Finally, scaling to multi-qubit batteries would illuminate whether the first-order transition persists or gives way to mean-field critical behavior, connecting our finite-system results to the extensive literature on many-body open quantum systems \cite{Buca2012note}.

Overall, our findings provide a unified physical picture of how coherence, detuning, and non-Markovianity can be harnessed to control energy storage and work extraction. They demonstrate that environmental engineering and coherent control, rather than merely mitigating decoherence, can be actively exploited to drive quantum systems across sharp thermodynamic boundaries. These insights offer practical guidelines for designing high-performance, noise-resilient quantum batteries and establish ergotropy phase transitions as a new frontier in quantum thermodynamics.

\bibliographystyle{ieeetr}
\bibliography{ref}
\end{document}